\documentclass[%
 amsmath,amssymb,
 aps,
]{revtex4-1}

\usepackage{graphicx}
\usepackage{dcolumn}
\usepackage{bm}

\begin{document}

\title{ Black hole as topological insulator (II): the boundary modes}
\author{Jingbo Wang}
\email{ shuijing@mail.bnu.edu.cn}
\affiliation{Institute for Gravitation and Astrophysics, College of Physics and Electronic Engineering, Xinyang Normal University, Xinyang, 464000, P. R. China}
 \date{\today}
\begin{abstract}
 In the previous paper Ref.\cite{wangti1}, it was claimed that the black hole can be considered as a kind of topological insulator. For BTZ black hole in three dimensional $AdS_3$ spacetime two evidences were given to support this claim: the first evidence comes from the  black hole ``membrane paradigm", and the second evidence comes from the fact that the horizon of BTZ black hole can support two chiral massless scalar field with opposite chirality. Those are two key properties of 2D topological insulator. For higher dimensional black hole the first evidence is still valid but the second fails. In this paper, starting from the boundary BF theory, which can be used to describe the boundary degrees of freedom of black hole in arbitrary dimension, we shown that the isolated horizon of $3+1-$D black hole can support massless scalar field and vector field. Those two fields can be used to construct a massless Dirac field through the $2+1-$dimensional bosonization, which also appears on the boundary of $3+1-$D topological insulators.
\end{abstract}
\maketitle
\section{Introduction}
Black hole has attract people's attention since a long time ago. The pioneering work of Bekenstein\cite{bek1}, Hawking\cite{hawk1} have suggested that black hole have temperature and entropy. So it is important to consider the quantum theory of black hole. But there are difficult problems with quantum black hole, such as information loss paradox\cite{info1}, the firewall paradox\cite{firewall1} and so on. Up to now, we still don't have a satisfying quantum theory of black hole.

In the previous paper Ref.\cite{wangti1}, it was claimed that the black hole can be considered as a kind of topological insulator. For BTZ black hole in three dimensional $AdS_3$ spacetime two evidences were given to support this claim: the first evidence comes from the  black hole ``membrane paradigm"\cite{mb2}, and the second evidence comes from the fact that the horizon of BTZ black hole can support two chiral massless scalar field with opposite chirality\cite{whcft1}. Those are two key properties of 2D topological insulator\cite{2d1,2d2,2d3}. For higher dimensional black hole the first evidence is still valid. Since the general relativity can't be rewritten as Chern-Simons theory in higher dimensional spacetime, the second evidence fails. In this paper, starting from the boundary BF theory\cite{hw1}, which can be used to describe the boundary degrees of freedom of black hole in arbitrary dimension\cite{wmz,wang1,wh1,wh2,wh3}, it shown that the isolated horizon of $3+1-$D black hole can support massless scalar field and vector field. Those two field can be used to construct a massless Dirac field through the $2+1-$dimensional bosonization\cite{boson1}, which appeared on the boundary of $3+1-$D topological insulators\cite{tibf1}.

The article is organized as follows. In section 2 we consider the boundary modes on the horizon of BTZ black hole. In section 3 the results are generalized to $3+1-$D black hole. Section 4 is the conclusion.
\section{BTZ black hole}
In this section, we study the simpler case: the BTZ black hole in three dimensional $AdS_3$ spacetime. In the previous paper\cite{whcft1,wangti1}, it was shown that the horizon can have two chiral massless scalar fields with opposite chirality. The starting point is the reformulation of gravity theory in terms of Chern-Simons theory\cite{at1,witten1}. But this method restrict to three dimension due to the property of Chern-Simons theory. On the other hand, it was shown that the degrees of freedom on the horizon can be described by BF theory with sources\cite{wmz,hw1}. This result hold for wider case, including general relativity in arbitrary dimension\cite{wang1,wh1}, Lovelock theory\cite{wh2} and scalar-tensor theory\cite{wh4}.

To study the field theory on the horizon of BTZ black hole, the advanced Eddington coordinate $(v,r,\varphi)$ is more suitable. The metric of BTZ black hole is
\begin{equation}\label{1}
    ds^2=-N^2 dv^2+2 dv dr+r^2 (d\varphi+N^\varphi dv)^2.
\end{equation}
Choose the following Newman-Penrose null co-triads
\begin{equation}\label{2}
    l=-\frac{1}{2}N^2 dv+dr,\quad n=-dv,\quad m=r N^\varphi dv+r d\varphi,
\end{equation}
the corresponding spin connection is $A_2=\alpha m -\kappa n$ with $\alpha=N^\varphi,\kappa=r/L^2-r(N^\varphi)^2$.

On the horizon $r=r_+$, we choose $(x^0,x^1)=(v,\varphi)$. The boundary BF theory on the horizon $\Delta$ is given by\cite{wang1}
\begin{equation}\label{3}
    S=\int_{\Delta}BF=\int_{\Delta}B {\rm d}A,
\end{equation}
with
\begin{equation}\label{4}
    {\rm d}B=\frac{1}{8\pi G}m,\quad A=\kappa dv+d\beta,
\end{equation}
where $A$ is the non-rotating component of the connection $A_2$, and $\beta$ represent the SO$(1,1)$ symmetry on the horizon. Note that this form of BF theory also appeared in the BF theory description of topological insulator \cite{tibf1} (the Eq.(42)).

Up to a boundary term, the action can be rewritten as
\begin{equation}\label{5}
    S=\frac{1}{2}\int_{\Delta}B {\rm d}A-{\rm d}B\wedge A,
\end{equation}
The canonical form is
\begin{equation}\label{6}\begin{split}
    S=\frac{1}{2}\int_{\Delta}B(\partial_0 A_1-\partial_1 A_0)-( A_1\partial_0 B-A_0\partial_1 B )\\
    =\frac{1}{2}\int_{\Delta}B\partial_0 A_1+A_0\partial_1 B -A_1\partial_0 B +A_0\partial_1 B .
\end{split}\end{equation}
Choose the gauge $A_0=0$ to gives
\begin{equation}\label{6}\begin{split}
    S=\frac{1}{2}\int_{\Delta}B\partial_0 A_1-A_1\partial_0 B.
\end{split}\end{equation}
From (\ref{4}) it can be shown that the gauge $A_0=0$ imply
\begin{equation}\label{7}
    \partial_0 \beta=-\kappa,\quad A_1=\partial_1 \beta.
\end{equation}
Insert into the action (\ref{6}) to give
\begin{equation}\label{8}\begin{split}
    S=\frac{1}{2}\int_{\Delta}B\partial_0 \partial_1 \beta-\partial_1 \beta\partial_0 B\\
    =-\frac{1}{2}\int_{\Delta}\partial_1 B\partial_0  \beta+\partial_1 \beta \partial_0 B
\end{split}\end{equation}
which is the same as the Eq.(49) in Ref.\cite{tibf1}. So we can get two chiral bosonic modes flowing in opposite directions. Those bosonic modes can combined into a one-dimensional fermionic system via standard $1+1-$dimensional bosonization.

Now we have two fields $B,\beta$. To construct a massless fermionic field those two fields should satisfy the dual relation\cite{boson1}
\begin{equation}\label{9}
    \partial_\mu \beta+\epsilon_{\mu \nu}\partial^\nu B=0.
\end{equation}
But due to the constraint (\ref{4}) and (\ref{7}), the dual relation don't hold. We must re-scale those fields
\begin{equation}\label{9a}
   \tilde{\beta}=\gamma_1 \beta, \tilde{B}=\gamma_2 B.
\end{equation}
We want the coefficient of BF action to be a dimensionless constant $\frac{1}{C}$, then
\begin{equation}\label{10}
 \gamma_1=\sqrt{\frac{C}{8\pi G \kappa}},\quad  \gamma_2=\pi \sqrt{8\pi C G \kappa}.
\end{equation}

The final action for BF theory on the horizon of BTZ black hole is
\begin{equation}\label{11}
    S=\frac{1}{C}\int_{\Delta}\tilde{B} {\rm d}\tilde{A},
\end{equation}
with $\tilde{A}=\sqrt{\frac{C}{8\pi G \kappa}} \bar{A}$. When $C=\pi$, one can get the BF theory for topological insulator \cite{tibf1}.
\section{Four dimensional black hole}
Unlike the Chern-Simons theory, the BF theory can be applied to 4D and higher dimensional black holes. Similar to the case in 3D, the boundary BF action on the isolated horizon is
\begin{equation}\label{12}
    S=\int_{\Delta}BF=\int_{\Delta}B \wedge {\rm d}A,
\end{equation}
with
\begin{equation}\label{14}
    {\rm d}B=\frac{\Sigma_{01}}{8\pi G}=\frac{1}{8\pi G} e^2\wedge e^3,\quad A=\kappa dv+d\beta.
\end{equation}
The action can also be written as
\begin{equation}\label{15}
    S=\frac{1}{2}\int_{\Delta}B\wedge {\rm d}A+{\rm d}B\wedge A,
\end{equation}
The canonical form is
\begin{equation}\label{16}\begin{split}
    S=\frac{1}{2}\int_{\Delta} \epsilon^{\mu\nu\rho}B_\mu F_{\nu\rho}+\epsilon^{\mu\nu\rho}\partial_\mu B_\nu A_\rho\\
    =\frac{1}{2}\int_{\Delta}2B_0(\partial_1 A_2-\partial_2 A_1)+2A_0(\partial_1 B_2-\partial_2 B_1) -A_1\partial_0 B_2 +A_2\partial_0 B_1-B_1\partial_0 A_2 +B_2\partial_0 A_1 .
\end{split}\end{equation}

The gauge $A_0=0$ imply
\begin{equation}\label{17}
    \partial_0 \beta=-\kappa,\quad A_i=\partial_i \beta\Rightarrow \partial_1 A_2-\partial_2 A_1=0.
\end{equation}
Insert into the action (\ref{16}) to give
\begin{equation}\label{8}\begin{split}
    S=\frac{1}{2}\int_{\Delta}B_2\partial_0 \partial_1 \beta-\partial_1 \beta\partial_0 B_2-B_1\partial_0 \partial_2 \beta+\partial_2 \beta\partial_0 B_1\\
    =-\frac{1}{2}\int_{\Delta}\partial_0  \beta\epsilon^{ij}\partial_i B_j+\partial_i \beta \epsilon^{ij} \partial_0 B_j,
\end{split}\end{equation}
which is the same as the Eq.(53) in Ref.\cite{tibf1}.

Now we have two fields $\beta$ and $B$. To construct a massless fermionic field those two fields should satisfy the dual condition\cite{boson1}
\begin{equation}\label{19}
    \partial_\mu \beta+\epsilon_{\mu \nu\rho}\partial^\nu B^\rho=0.
\end{equation}
First taking $\mu=0$ to gives
\begin{equation}\label{20}
    \kappa=-\partial_0 \beta=\epsilon_{ij}\partial^i B^j=\frac{1}{8\pi G}.
\end{equation}
Due to the zero law of isolated horizon, the surface gravity $\kappa$ is a constant on the horizon. Similar to the BTZ case, we can re-scale the two fields to satisfying the above condition and fixed coefficient. The result is
\begin{equation}\label{21}
   \tilde{\beta}=\sqrt{\frac{C}{8\pi G \kappa}} \beta, \quad \tilde{B}=\pi \sqrt{8\pi C G \kappa} B.
\end{equation}
Let's consider the canonical mass dimension of those fields. The spin connection has dimension $[A_\mu]=1$, the tetrad $[e_\mu]=0$, so one can get $[B_\mu]=1,[\beta]=0$. Since $[\kappa]=1,[G]=-2$, so $[\tilde{\beta}]=[\tilde{B}]=\frac{1}{2}$, which are the correct canonical mass dimension for scalar field and vector field in 3 dimension.

Next taking $u=i$ to gives
\begin{equation}\label{22}
    \partial_i \beta=\epsilon_{ij}\partial^0 B^j.
\end{equation}
Choose gauge $B^0=0$, then the above equation gives the Coulomb gauge condition $\partial_i B^i=0$.

The final action for BF theory is
\begin{equation}\label{23}
    S=\frac{1}{C}\int_{\Delta}\tilde{B}\wedge {\rm d}\tilde{A},
\end{equation}
with $\tilde{A}=\sqrt{\frac{C}{8\pi G \kappa}} A$.
\section{Conclusion}
In this paper, we study the boundary field theory for the BTZ black hole and 4D black hole from the BF theory. For BTZ black hole, the boundary theory contain two chiral bosonic modes flowing in opposite direction. Through standard $1+1-$dimensional bosonization, those bosonic modes can combined into a one-dimensional fermionic system. For black hole in four dimensional spacetime, the BF theory transforms into a theory that contain a scalar field and a vector field, which can construct massless fermionic fields through $2+1-$dimensional bosonization. Those fermionic field also appears in the boundary of topological insulators.

But we don't need to stop here. The BF theory can describe the degrees of freedom on the horizon for arbitrary dimensional black holes. Consider the $n+1-$dimensional spacetime. The horizon is $n-$dimension, and the BF theory reduces to a theory that contain a scalar field $\beta$ and a $(n-2)-$rank form field $B_{\mu\nu\ldots \rho}$. This $n-2-$rank form field $B$ is crucial for higher-dimensional bosonization\cite{highbo1}. For example, in $4+1-$dimensional spacetime, those scalar-tensor fields can combined into a Weyl spinor in $3+1-$dimensional boundary\cite{highbo2}.

Due to the periodic table of topological insulators and superconductors \cite{pt1,pt2,pt3}, in every dimension, there are five types of non-trivial topological phases, two indexing by $Z_2$ and three by $Z$. The boundary modes of topological insulator in other dimension(not 2 and 3) may not be protected by time reversal symmetry, but by other symmetry. In black hole case, the boundary modes are protected not by symmetry, but by geometry. On the isolated horizon of black hole, due to the properties of isolated horizon, there exist the massless boundary modes. So the boundary modes are protected by the geometry of isolated horizon, and are difficult to destroy. Also the horizon of the black hole, which can conduct electricity, is intrinsic to black hole, so the black hole is more like the topological insulator rather than an insulator with a sheet of aluminum foil on it.

The coefficient of BF theory is important, since for $C=\pi$ it describe the integral topological insulator, and for other values it may describe the fractional topological insulators \cite{tibf1}. We have not find a canonical way to fix the value of $C$ from the black hole physics, so it is not clear that the black hole is integral or fractional topological insulators.

\acknowledgments
 This work is supported by the NSFC (Grant No.11647064) and Nanhu Scholars Program for Young Scholars of XYNU.
 \bibliography{it2}
\end{document}